\documentclass[prl,superscriptaddress,reprint,showpacs,twocolumn,balance,notitlepage]{revtex4-1}

\usepackage{graphicx} %[dvips]
\usepackage{tikz}
\usepackage{amsmath}
\usepackage{amssymb}
\usepackage{color}
\usepackage{float}
\usepackage{epstopdf}
\usepackage[normalem]{ulem}
\usepackage{pgfplots,pgfplotstable}
\usepackage{soul}
\usepackage{accents}
\usepackage{nameref}
\usepackage{balance}
\usepackage[all]{nowidow}

\makeatletter
\newcommand*{\balancecolsandclearpage}{%
  \close@column@grid
  \clearpage
  \twocolumngrid
}
\makeatother

\usetikzlibrary[shadings]
\usetikzlibrary{shapes}
\usetikzlibrary{snakes}
\usepgfplotslibrary{fillbetween}
\tikzset{
    axis break gap/.initial=10mm
}
\usetikzlibrary{patterns}

\pgfplotsset{compat=newest}

\sethlcolor{white}

\newcounter{marknumber}
\pgfplotsset{
    error bars/every nth mark/.style={
        /pgfplots/error bars/draw error bar/.prefix code={
            \pgfmathtruncatemacro\marknumbercheck{mod(floor(\themarknumber/2),#1)}
            \ifnum\marknumbercheck=0
            \else
                \begin{scope}[opacity=0]
            \fi
        },
        /pgfplots/error bars/draw error bar/.append code={
            \ifnum\marknumbercheck=0
            \else
                \end{scope}
            \fi
            \stepcounter{marknumber}    
        }
    }
}

\newcommand{\bs}{\boldsymbol}
\newcommand{\vett}[1]{\mathbf{#1}}

\newcommand {\tr} {\mbox{\rm tr\,}}

\newcommand {\dd}{\mathrm{d}}
{\left\lbrace\begin{array}{@{}l@{}}}%
{\end{array}\right.}

\begin{document}

\title{Curvature-Induced Instabilities of Shells}

\author{Matteo Pezzulla}
\affiliation{
Department of Mechanical Engineering, Boston University, Boston, MA, 02215.
}%

\author{Norbert Stoop}
\affiliation{
Department of Mathematics, Massachusetts Institute of Technology - Cambridge, MA, 02139.
}%

\author{Mark P. Steranka}
\affiliation{
Department of Mechanical Engineering, Boston University, Boston, MA, 02215.
}%

\author{Abdikhalaq J. Bade}
\affiliation{
Department of Mechanical Engineering, Boston University, Boston, MA, 02215.
}%

\author{Douglas P. Holmes}
\email{dpholmes@bu.edu}
\affiliation{
Department of Mechanical Engineering, Boston University, Boston, MA, 02215.
}%

\date{\today}

\begin{abstract}
Induced by proteins within the cell membrane or by differential growth, heating, or swelling, spontaneous curvatures can drastically affect the morphology of thin bodies and induce mechanical instabilities. Yet, the interaction of spontaneous curvature and geometric frustration in curved shells remains poorly understood. Via a combination of precision experiments on elastomeric spherical shells, simulations, and theory, we show how a spontaneous curvature induces a rotational symmetry-breaking buckling as well as a snapping instability reminiscent of the Venus fly trap closure mechanism. The instabilities, and their dependence on geometry, are rationalized by reducing the spontaneous curvature to an effective mechanical load. This formulation reveals a combined pressurelike term in the bulk and a torquelike term in the boundary, allowing scaling predictions for the instabilities that are in excellent agreement with experiments and simulations. Moreover, the effective pressure analogy suggests a curvature-induced subcritical buckling in closed shells. We determine the critical buckling curvature via a linear stability analysis that accounts for the combination of residual membrane and bending stresses. The prominent role of geometry in our findings suggests the applicability of the results over a wide range of scales.
\end{abstract}


\maketitle

Owing to their slender geometry, thin elastic shells display intriguing mechanical instabilities. Perhaps the most iconic example is the buckling of a spherical shell under pressure - a catastrophic situation that often leads to structural failure~\cite{Bushnell,Koiter1969}. Instabilities and shape changes are also fundamental during the development and morphogenesis of thin tissue~\cite{Katifori2010, Lim2002}. To control and evolve shape, Nature heavily relies on internal stimuli such as growth, swelling, or active stresses~\cite{Liang2011, Tallinen2014}. If the stimulus varies through the thickness of the shell, it generally induces a change of the spontaneous (or natural) curvature of the tissue~\cite{Goriely2017}. Examples are the ventral furrow formation in Drosophila~\cite{Heer2017} or the fast closure mechanism invoked by the Venus fly trap to catch prey~\cite{Forterre2005}. Harnessing similar concepts for technological applications, internal stimuli were also suggested as a means to design adaptive metamaterials~\cite{Holmes2007} and soft robotics actuators~\cite{Yuk2017}. To describe the mechanics of slender structures with arbitrary stimuli, classical shell mechanics was extended recently to model bodies that do not possess a stress-free configuration~\cite{Gurtin2010, AmarGoriely2005,Goriely2005}, leading to the non-Euclidean shell theory~\cite{Efrati2009}. Despite recent progress~\cite{Armon2011,Pezzulla2016}, the role of curvature-altering stimuli, and their interplay with geometric frustration and instabilities in thin, initially curved shells, remains poorly understood.

In this Letter, we combine precision experiments with non-Euclidean shell theory to reveal how curvature stimuli induce mechanical instabilities in spherical shells. Our experiments demonstrate symmetry-breaking as well as snap-through shape transitions depending on the amount of stimulus and the deepness of the shell. To rationalize our findings, we show that a curvature stimulus reduces to a pressure-like normal force in the bulk, but induces a torque along the boundary of the shell. A scaling analysis of the dominant boundary term allows us to construct an analytical phase diagram that captures well the transitions found in experiments and simulations. For closed spherical shells, we show that the pressure-like stimulus induces a curvature-controlled buckling instability. The critical stimulus is obtained from stability analysis and found to be in the range of related biological systems. 

\begin{figure}[t]
\centering
\hspace{2pt}
\includegraphics[scale=1]{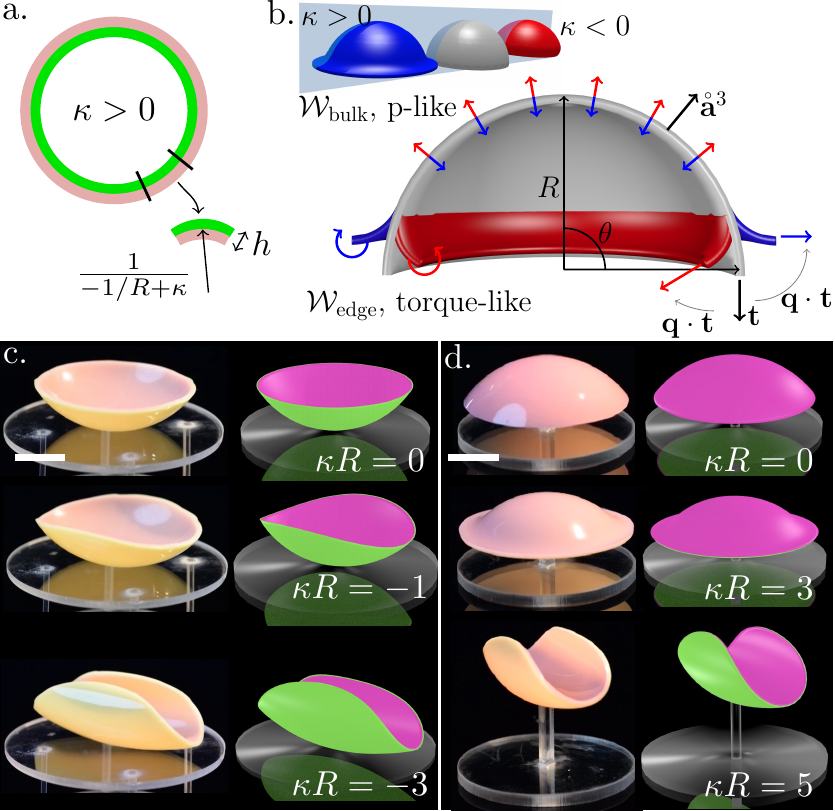}
\caption{(a) Schematic of a VPS bilayer shell with natural radius of curvature~$1/(-1/R+\kappa)$ induced by residual swelling. (b) The natural curvature mechanically corresponds to torques on the boundary and a pressure field in the bulk. (c) Buckling of open spherical shells triggered by~$\kappa<0$, (left: experiments; right: simulations). (d) Snapping of open spherical shells triggered by~$\kappa>0$ for~$\bar{\theta}=5$. Scale bars~$2$~cm.\label{instability}}
\end{figure}

In our experiments, we uniformly coated a rigid sphere (radii~$R\in[12,75]$~mm) with silicone-based vinyl-polysiloxane (VPS) 32 (Zhermack), such that it thermally crosslinks into an elastomeric shell~\cite{Lee2016}. We then repeated the coating process with VPS 8, and cut shells with opening angles~$\theta\in[20,150]^\circ$, resulting in bilayer shells of thicknesses~$h\in[0.5,1.3]$~mm. Due to differential swelling between the two polymer layers, internal stresses develop. We quantify this geometric frustration by cutting a long, narrow strip from the shell. Free of any constraints, the strip adopts a shape with curvature~$\bar{\kappa}$, which can be additively decomposed into the initial curvature~$-1/R$ and natural curvature~$\kappa$. Thus, $\kappa = \bar{\kappa} + 1/R$ measures the curvature stimulus (Fig.~\ref{instability}~(a))~\cite{Pezzulla2015,Pezzulla2016}. Specifically, for a bilayer with VPS 8 on the outside, we find $\kappa>0$, and by switching the order of the layers, we can induce a negative natural curvature ($\kappa<0$). To characterize the various geometries, we introduce the dimensionless parameter~$\bar{\theta}=\theta/\sqrt{h/R}$, describing the \emph{deepness} of the shell with respect to the angular width of the boundary layer~$\sqrt{h/R}$~\cite{Niordson1985}.

\begin{figure}[tp]
\hspace{-20.5pt}
\includegraphics[scale=1]{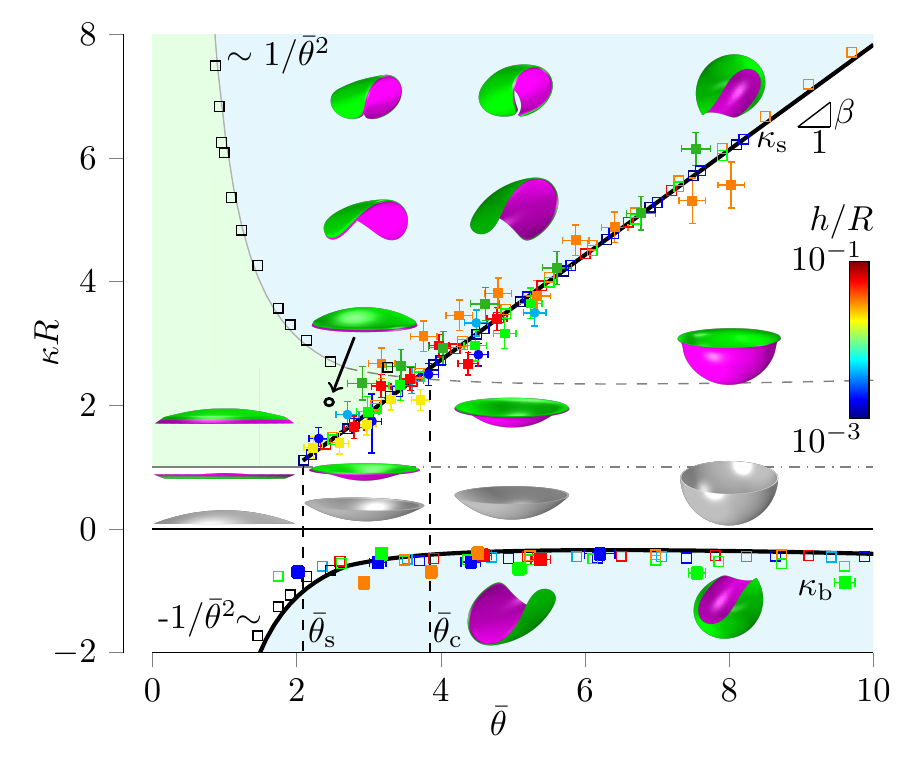}
\hspace{-20.5pt}
\vspace{-0.5cm}
\caption{Phase diagram of curvature-induced instabilities in open shells: white and green regions denote phases with rotational symmetry but opposite surface orientations, whereas blue regions denote phases of broken rotational symmetry. Theoretical transitions curves (solid lines) match well with experimental (colored full symbols) and numerical (colored empty symbols) findings, where color represents~$h/R$. \label{phasediagram}}
\end{figure}

For shells with~$\kappa<0$, we find that the stimulus leads to a loss of rotational symmetry via a supercritical buckling bifurcation (Fig.~\ref{instability}~(c))~\cite{Katifori2010}. Experiments suggest no strong dependence of this transition on~$\bar{\theta}$. For~$\kappa>0$, the stimulus acts to evert the initial curvature of the shell. Above a critical stimulus, we observe a snap-through instability (Fig.~\ref{instability}~(d)), reminiscent of the abrupt concave-convex shape changes employed by the Venus fly trap~\cite{Forterre2005}, and the embryonic inversion of \emph{Volvox}~\cite{Hohn2015}. Here, the critical curvature stimulus increases with~$\bar{\theta}$. Moreover, shallow shells with~$\bar{\theta}<\bar{\theta}_\textup{s}\approx 2$ do not snap, whereas shells with~$\bar{\theta}<\bar{\theta}_\textup{c}\approx 4$ remain rotationally symmetric after snapping, while deep ones break the rotational symmetry during snap-through (Fig.~\ref{phasediagram}). 

To explain the richness of the experimental findings, we rely on non-Euclidean shell theory that has recently been proposed as a model for growth in thin, bidimensional bodies~\cite{Efrati2009}. In this formulation, the mechanics of the shell is entirely described by the geometry of the middle-surface with first and second fundamental forms~$\vett{a}, \vett{b}$~\cite{Oneill1997}. The undeformed \textit{reference configuration} in absence of curvature stimulus is characterized by~$\accentset{\circ}{\vett{a}}$, $\accentset{\circ}{\vett{b}}$, respectively. Curvature stimuli are modeled by changing the reference configuration to effectively generate stresses and moments arising from differential swelling of the shell layers~\cite{Pezzulla2017}. The resulting \textit{natural configuration} has fundamental forms~$\bar{\vett{a}}$, $\bar{\vett{b}}$ and is generally not embeddable in Euclidean space. When the stimulus does not induce a stretch of the mid-surface ($\bar{\vett{a}}=\accentset{\circ}{\vett{a}}$), one obtains
$\bar{\vett{b}}=\accentset{\circ}{\vett{b}}+\kappa\accentset{\circ}{\vett{a}}$, where~$\kappa$ is the scalar (additive) natural curvature~\cite{Pezzulla2017}.
The energy of the shell may be written after some algebra as~\footnote{See supplementary information for a detailed derivation, which includes Refs.~\cite{Deserno2004,Cirak2001,Neut1932,heijden2008}}
\begin{equation}\label{energy}
\overline{\mathcal{U}}=\overline{\mathcal{U}}_{\textup{s}}^{\textup{K}}+\frac{h^2}{3}\overline{\mathcal{U}}_{\textup{b}}^{\textup{K}}-\frac{2(1+\nu)h^2}{3}\int\kappa\ \tr({\vett{b}}-\accentset{\circ}{\vett{b}})\,\dd\omega\,.
\end{equation}
Here, $\overline{\mathcal{U}}^\textup{K}=\overline{\mathcal{U}}_{\textup{s}}^{\textup{K}}+h^2\overline{\mathcal{U}}_{\textup{b}}^{\textup{K}}/3$ is Koiter's classical shell energy composed of stretching and bending terms without any inelastic stimuli~\cite{Note1,Koiter1973}, $\nu$ is the Poisson ratio ($\nu=1/2$ for VPS), and~$\dd\omega$ is the area element~\cite{Hanna2017}. Owing to the additive decomposition, we can interpret the last term in~\eqref{energy} as the stimulus-induced curvature potential~$\mathcal{P}_\kappa=-2(1+\nu)h^2/3\int\kappa\ \tr({\vett{b}}-\accentset{\circ}{\vett{b}})\,\dd\omega$. The surprisingly simple additive effect of natural curvature allows for a relatively straightforward extension of thin shells simulation methods to minimize~\eqref{energy} for a given stimulus~$\kappa$. Indeed, by numerically minimizing Eq.~\eqref{energy}, we find good quantitative agreement with the experimental shapes and the stimulus-induced transitions (Fig.~\ref{instability}~(c, d)). This suggests that the reduced-order model~\eqref{energy} is adequate to describe thin shells with curvature stimuli.

To theoretically understand how natural curvature interacts with the geometry and triggers the observed instabilities, we analyze the curvature potential and provide its geometrical interpretation. We start by expanding~$\tr(\vett{b}-{\accentset{\circ}{\vett{b}}})$ in terms of the displacement field~$\vett{\Psi}$ up to first order~\cite{Niordson1985}. Assuming a homogeneous natural curvature stimulus~$\kappa$, the curvature potential decouples into bulk and boundary terms, $\mathcal{P}_\kappa=-\mathcal{W}_\textup{bulk}-\mathcal{W}_\textup{edge}$~\cite{Note1}. For a sphere with outward pointing normal, they read
\begin{subequations}
\begin{eqnarray}\label{bulkwork}
\mathcal{W}_\textup{bulk}&=&-\frac{4(1+\nu)}{3}\Bigl(\frac{h}{R}\Bigr)^2\kappa\int \Psi_3\,\dd\omega\,,\\ \label{boundarywork}
\mathcal{W}_\textup{edge}&=&\frac{2(1+\nu)}{3}h^2\kappa\oint\Bigl(\vett{q}-\frac{\check{\vett{\Psi}}}{R}\Bigr)\cdot\vett{t}\,\dd s\,,
\end{eqnarray}
\end{subequations}
where~$\Psi_3$ is the normal displacement, $\check{\vett{\Psi}}$ is the in-plane displacement field, and~$\vett{t}$ is the outward normal vector to the boundary curve. $\vett{q}=\nabla\Psi_3-\check{\vett{\Psi}}/R$ represents the rotation of an element of the shell~\cite{Niordson1985}, such that~$\vett{q}\cdot\vett{t}$ is the rotation of~$\vett{t}$ (Fig.~\ref{instability}~(b)). The integral in~\eqref{bulkwork} is equivalent to the first-order energy of a pressure load. In the bulk, a curvature stimulus is therefore equivalent to an effective applied pressure. In~\eqref{boundarywork}, $\kappa$ is the work conjugate of the rotation~$\vett{q}\cdot\vett{t}$ and the membrane in-plane displacement $\check{\vett{\Psi}}\cdot\vett{t}$, implying both a torquelike and membrane force-like behavior. Specifically, for $\kappa>0$, Eq.~\eqref{boundarywork} describes an outward torque at the boundary (Fig.~\ref{instability}~(b)). A similar interpretation holds for arbitrary open shells~\cite{Note1}.

Numerically, we find that $|\mathcal{W}_\textup{edge}| \gg |\mathcal{W}_\textup{bulk}|$ for thin shells of all considered opening angles $\theta$. We can rationalize this by considering small displacements. The Koiter elastic energy then scales as~$\overline{\mathcal{U}}^{\textup{K}}\sim(\Psi_3/R)^2R^2$, while the curvature potential scales as~$\mathcal{P}_\kappa\sim h^2\kappa(\Psi_3/R^2)R^2$ in the bulk. A balance of the two leads to~$\Psi_3\sim h^2\kappa$. As the area of the shell is proportional to~$R^2(1-\cos\theta)$, the bulk work~\eqref{bulkwork} scales as~$\mathcal{W}_\textup{bulk}\sim h^4\kappa^2(1-\cos\theta)$. Then, as the boundary layer is bending dominated~\cite{Efrati2009a}, we obtain~$|\vett{q}-\check{\vett{\Psi}}/R|\sim\kappa\sqrt{Rh}$~\cite{Note1}, where~$\sqrt{Rh}$ is the characteristic width of the boundary layer~\cite{Niordson1985}. As the perimeter of the boundary is proportional to~$R\sin\theta$, we conclude that the edge work~\eqref{boundarywork} scales as~$\mathcal{W}_\textup{edge}\sim h^4\kappa^2(R/h)^{3/2}\sin\theta$. By a comparison of the two scalings, we find~$|\mathcal{W}_\textup{edge}/\mathcal{W}_\textup{bulk}|\sim(R/h)^{3/2}/\tan(\theta/2)\gg 1$, i.e. the boundary work dominates for the opening angles $\theta$ considered. Therefore, the boundary term dictates the observed shape transitions.

In experiments, we observe that snapping is indeed accompanied by minimal bulk deformation, but large rotation of the boundary. Moreover, we find that snap-through instabilities occur for open shells with~$\theta\le\pi/2$ when their tangent plane on the boundary becomes approximately horizontal (see the supplementary videos). In this state, the critical curvature within the boundary layer scales as~$b_\textup{c} \sim (1+\nu)(-1/R+\kappa)$~\cite{Note1}. Since the width of the boundary layer scales as~$\sqrt{Rh}$, $b_\textup{c}$ must also scale as $\sim \theta/\sqrt{Rh}$. Thus we find that the critical curvature stimulus at snapping~$\kappa_\textup{s}R\sim\bar{\theta}$, that is
\begin{equation}\label{scaling_snapping}
\kappa_\textup{s}R=\beta\bar{\theta}-\alpha\,,
\end{equation}
leaving two scaling coefficients~$\alpha$ and~$\beta$ to be determined later. 
For~$\bar{\theta}\rightarrow0$, shells tend to plates. Flat plates of radius~$r$ under curvature stimuli bifurcate at~$\tilde{\kappa}_\textup{p}h=\pm a(h/r)^2$ with~$a=\sqrt{10+7\sqrt{2}}$~\cite{Pezzulla2016}. Then, for large~$R$ and small~$\theta$, but~$r=R\theta$ finite, shells are expected to behave like plates if we identify~$\tilde{\kappa}_\textup{p}R=\kappa_\textup{p} R-1$, \textit{i.e.} we compensate for the initial curvature~$-1/R$. Therefore, shells will bifurcate at~$\kappa_\textup{p}R=\pm a/\bar{\theta}^2+1$, and we expect a symmetric bifurcation behavior around~$\kappa R=1$. Without loss of generality, we consider the case~$\kappa<0$, corresponding to the buckling of shells into spindle-like shapes. We define the critical curvature stimulus by~$\kappa_{\textup{b}}$, and now consider the behavior of deep shells. We note that for~$\theta\rightarrow\pi$, the natural curvature will expend a torquelike work on a boundary whose perimeter approaches zero as~$\sin\theta$, while the area of the shell to be deformed increases as~$(1-\cos\theta)$. The critical natural curvature will then diverge as~$\kappa_\textup{b}R\sim\tan(\theta/2)\sim1/(\theta-\pi)$, that is~$1/(\bar{\theta}-\pi\sqrt{R/h})$. We conjecture that the curvature buckling of shells can be determined by combining the two diverging regimes for small and large~$\bar{\theta}$ as
\begin{equation}\label{buckling}
\kappa_\textup{b}R=-\frac{a}{\bar{\theta}^2}+1+\frac{b}{\bar{\theta}-\pi\sqrt{R/h}}+c\,,
\end{equation}
where~$a$ was given above, and~$b$ and~$c$ have to be determined by fitting to simulations. Notice that the superposition of the two scalings retains the correct asymptotic behaviors as~$\bar{\theta}\rightarrow0$ and~$\bar{\theta}\rightarrow\pi\sqrt{R/h}$. 

Our theoretical scaling predictions can be summarized in a phase diagram (solid lines in Fig.~\ref{phasediagram}) in the parameters $(\bar{\theta},\kappa R)$, which fully characterize the curvature-induced instabilities of open shells. For~$\kappa<0$, the scaling law~\eqref{buckling} with~$b=3.6$ and~$c=-0.98$ provides the best fit with numerics, and agrees well with experiments. We note that a parameter-free determination of the buckling threshold would require a linear stability analysis, which is hampered due to the nontrivial fundamental state before buckling. For~$\kappa>0$, the behavior is richer: there are two phases of inverted curvature, one with broken rotational symmetry (blue region), and another phase that is rotationally symmetric (green). Simulations confirm the snapping transition~\eqref{scaling_snapping} with~$\alpha=0.67$ and~$\beta=0.85$, but only if~$\bar{\theta}>\bar{\theta}_\textup{s}=2.09$, in agreement with experiments. For~$\bar{\theta}<\bar{\theta}_\textup{s}$, we find that shells smoothly invert their curvature into the green phase as~$\kappa$ increases. This can be understood by considering a family of shells with a fixed~$h/R$, and different~$\theta$. Since the width of the boundary layer scales as~$\sqrt{Rh}$, shallower shells possess a boundary layer that covers a larger portion of the area compared to deeper shells. Thus, there exists a critical value~$\bar{\theta}_\textup{s}$ below which the boundary layer transitions into a wide region that covers the entire shell. As regions within the boundary layer are bending-dominated, the curvature of the shell smoothly follows the evolution of the spontaneous curvature for~$\bar{\theta}<\bar{\theta}_\textup{s}$. Starting from the green phase, rotational symmetry is eventually lost for large~$\kappa R$. The transition line is well described by mirroring~Eq.~\eqref{buckling} around the axis of symmetry~$\kappa R=1$, as expected from the plate limit (dashed gray line), without any changes of the parameters~$b$ and~$c$. At~$\bar{\theta}_\textup{c}$, this transition line intersects with~Eq.~\eqref{scaling_snapping}, and a triple point emerges. Explicitly, the triple point is determined from~$-\kappa_\textup{s}+2/R=\kappa_\textup{b}$, which gives~$\bar{\theta}_\textup{c}=3.85$, in agreement with experiments ($\bar{\theta}_\textup{c}=3.95\pm0.26$). Consequently, shells snap into a rotationally symmetric phase only if~$\bar{\theta}_\textup{s}<\bar{\theta}<\bar{\theta}_\textup{c}$, whereas for~$\bar{\theta}>\bar{\theta}_\textup{c}$, shells immediately snap into an everted state of broken rotational symmetry (blue region; thin shells are unlikely to snap into cylindrical shapes~\cite{Abdullah2016}. However, we would expect the deformed shells have a small, nonzero curvature along one principal direction, corresponding to a near isometric deformation with minimum energy).  

In contrast to open shells, only the bulk contribution remains for closed shells. Exploiting its analogy with pressure, we expect instabilities similar to the classical pressure-induced buckling of spherical shells~\cite{VonKarman1939,Tsien1942,Thompson1964,Hutchinson1967,Koiter1973}. More precisely, the bulk term is formally equivalent to the work done by an external (dead) pressure, $\mathcal{W}_\textup{p}=-8(1-\nu^2)p/(Eh)\int\Psi_3\,\dd\omega$~\cite{Koiter1969}, allowing us to define an effective stimulus-induced pressure $p$ via 
\begin{equation}\label{analogy}
\kappa h=6(1-\nu)\Bigl(\frac{R}{h}\Bigr)^2\frac{p}{E}\,,
\end{equation}
where $E$ is the Young's modulus of the shell. Following this interpretation, a negative stimulus, $\kappa<0$, corresponds to a negative external pressure, $p<0$, thus causing an inflation of the shell. Conversely, a positive stimulus with~$\kappa>0$ is equivalent to positive external pressure and results in a compression of the sphere.  By expanding the bending and stretching strains up to the first order in the displacement~\cite{Niordson1985}, and solving the Euler-Lagrange equations associated to~\eqref{energy}, we find for the normal displacement 
$\Psi_3/h=-\kappa h/12+O((h/R)^4)$
while the in-plane displacement is zero for symmetry. Having established the analogy to classical shell buckling~\cite{Zoelly1915,Koiter1945}, we expect a critical stimulus beyond which the shell will buckle in absence of any external load. It is tempting to identify the buckling natural curvature~$\kappa_\textup{b}$ via~\eqref{analogy} with the critical buckling pressure~$p_\textup{b}=2E/\sqrt{3(1-\nu^2)}(h/R)^2$ obtained for the classical pressure buckling of spherical shells~\cite{Zoelly1915}. However, despite the formal analogy, pressure buckling and curvature buckling are triggered by fundamentally different residual stress states: while the residual stress in pressure-compressed shells is mainly of the membrane (in-plane) type, the pre-stress in curvature-compressed shells is a combination of membrane and bending stresses due to the evolving natural curvature that modifies the rest lengths of the body above and below the mid-surface. A careful analysis then yields the critical buckling stimulus as
\begin{equation}\label{kappabuckling}
\kappa_\textup{b}h=4\sqrt{3\frac{1-\nu}{(1+\nu)(5+4\nu)}}\,,
\end{equation}
which for an incompressible material reduces to~$\kappa_\textup{b}h=4/\sqrt{7}$~\cite{Note1}. This is a large value, corresponding to a radius of natural curvature equal to two-third's of the thickness (via residual swelling we are experimentally limited to values of~$|\kappa h|<1/4$), yet it is comparable to natural curvatures observed during the eversion of the \emph{Volvox} for which~$\kappa h\simeq2$~\cite{Hohn2015}.
\begin{figure}[tp] 
\hspace{-21pt}
\includegraphics[scale=1]{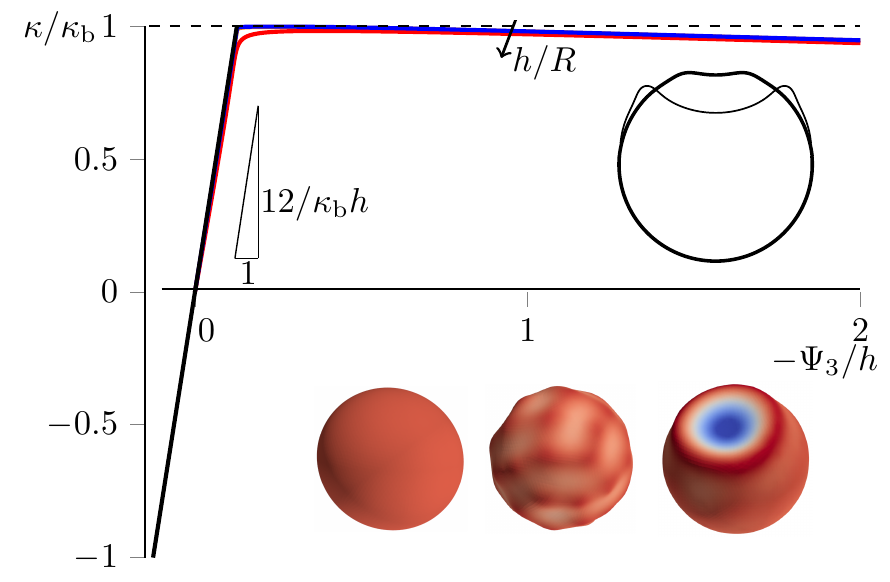}
\hspace{-21pt}
\caption{Curvature buckling of a closed shell. As the stimulus~$\kappa/\kappa_\textup{b}$ increases, the normal displacement at the north pole~$\Psi_3/h$ decreases linearly as predicted by theory. At~$\kappa=\kappa_\textup{b}$, buckling occurs and the shell becomes unstable. The solid black line represents theory, while the solid blue and red lines represent simulations for~$h/R=0.001,0.1$, respectively. Axisymmetric profiles and $3$D shapes from simulations are shown in the insets. \label{Psi3}}
\end{figure}
In contrast to open shells, where the characteristic curvature for snapping and buckling is~$1/R$ due to the existence of nearly isometric deformations, the characteristic curvature in closed shells becomes~$1/h$. To validate the buckling threshold, we performed simulations to minimize Eq.~\ref{energy} using closed spheres for different values of thicknesses and radii such that~$h/R\in[0.001,0.1]$. Measuring~$\Psi_3/h$ as we vary~$\kappa$, we confirm the behavior of~$\Psi_3/h=-\kappa h/12$ before buckling~\footnote{It is an open question to test whether a neo-hookean material will be affected by a limit point inflation instabilities triggered by~$\kappa<0$.}, as well as the predicted critical curvature~$\kappa_\textup{b}h$ in Eq.~\ref{kappabuckling} (Fig.~\ref{Psi3}). 
We note that after buckling, the shell becomes unstable as the bifurcation is subcritical. To track the lowest-energy unstable branch in~Fig.~\ref{Psi3}, we therefore minimized~Eq.~\ref{energy} using arc-length continuation while enforcing rotational symmetry (solid blue and red lines in Fig.~\ref{Psi3}). The post-buckling regime does not vary considerably with~$h/R$ and is similar to that observed in the pressure buckling of shells (insets in Fig.~\ref{Psi3})~\cite{Hutchinson2016}. 

In summary, we presented a theoretical and experimental study of curvature-induced instabilities in open and closed shells. Our theoretical analysis reveals that natural curvature can be interpreted as a combination of pressure and torque, and enables analytical predictions for instabilities in open and closed shells. We note that the critical stimuli in open and closed shells could also be determined via the method for nonlinear deformations presented in~\cite{AmarGoriely2005}, and a formal comparison  should be investigated in future work. We believe our study is a valuable contribution towards the generic understanding of curvature-driven instabilities in thin curved shells, as it generalizes previous experiments on plates~\cite{Pezzulla2016} and elastica with a natural curvature~\cite{Audoly2010}. Due to current limitations of the coating setup~\cite{Lee2016}, we hope that our study will motivate experiments on more general surfaces, e.g. via the application of advanced 3D printing techniques~\cite{Mao2016}. For simple geometries, the presented experimental setup is extensible to nonhomogenous stimuli by local patterning of the individual layers. We hypothesize that in the bulk, such stimuli remain at lowest order equivalent to normal forces, simplifying future theoretical analysis considerably. Lastly, the demonstrated precise control of shapes by means of natural curvature stimuli is scale-invariant, and thus presents novel means towards the design of self-folding and deployable structures as well as instability-driven actuators in soft robotics applications across different length scales.

%\subsection*{Acknowledgments}
We thank Miguel Trejo for help with preliminary experiments. We are also grateful to Jos\'e Bico and Beno\^{i}t Roman for helpful discussions. D.P.H. is grateful for financial support from the NSF CMMI--1505125.

%\subsection*{Author contributions}
%M.P. and D.P.H. designed research. M.P., M.P.S., and A.J.B. performed the experiments. M.P. developed the model, performed the calculations, and discussed the results with N.S. and D.P.H.. N.S. performed the 2D numerical simulations. M.P. performed the 1D and 3D numerical simulations. M.P., N.S., and D.P.H. wrote the paper.

\bibliography{biblio}

\balancecolsandclearpage
\begin{center}
\textbf{\large Supporting information}
\end{center}
%%%%%%%%%% Merge with supplemental materials %%%%%%%%%%
%%%%%%%%%% Prefix a "S" to all equations, figures, tables and reset the counter %%%%%%%%%%
\setcounter{equation}{0}
\setcounter{figure}{0}
\setcounter{table}{0}
\setcounter{page}{1}
\makeatletter
\renewcommand{\theequation}{S\arabic{equation}}
\renewcommand{\thefigure}{S\arabic{figure}}

\section*{Definitions}

For the following, we revise some basic geometric concepts to describe two-dimensional (2D) surfaces. Let $\mathcal{S}=\vett{r}(\eta^1, \eta^2)$ be such a 2D surface embedded in $\mathbb{R}^3$ and parameterized by $y=(\eta^1, \eta^2)$. We adopt the standard notation that Greek indices $\alpha,\beta,\ldots$ take values in $\{1,2\}$, whereas Latin indices $i,j,\ldots$  run from $1$ to $3$. The parametrization of $\mathcal{S}$ allows us to define (covariant) tangent vectors as
\begin{align}
\bs a_{\alpha} &= \vett{r}_{,\alpha} \equiv \frac{\partial \vett{r}}{\partial \eta^{\alpha}}\,,
\end{align}
as well as the induced surface metric $a_{\alpha \beta}$ (first fundamental form) 
\begin{align}\label{eq:first_fundamental_form}
a_{\alpha \beta} &= \bs a_{\alpha} \cdot \bs a_{\beta} = a_{\beta \alpha},
\end{align} 
where $\cdot$ denotes the Euclidean inner product on $\mathbb{R}^3$. The inverse metric is defined via $a^{\alpha \gamma} a_{\gamma \beta} = \delta^{\alpha}_{\beta}$, where we sum over repeated indices, and $\delta^{\alpha}_{\beta}$ denotes the Kronecker delta. The metric and its inverse map between co- and contravariant quantities, e.g. the contravariant form of $\bs a_1$ is
\begin{align}
\bs a^1 &= a^{1\alpha} \bs{a}_{\alpha}\,.
\end{align}
The  unit-length normal vector $\bs a_3$ is defined by
\begin{align}
\bs a_3 &\equiv \bs a^3= \frac{\bs a_1 \times \bs a_2}{|\bs a_1 \times \bs a_2|}\,.
\end{align}
Any vectorial surface quantity $\bs \Psi \in \mathbb{R}^3$, e.g. a displacement field, can thus be expressed as $\bs \Psi = \Psi^i \bs a_i = \Psi_i \bs a^i$, $i=1,2,3$. The surface element is 
\begin{equation}
\dd\omega = %\sqrt{a}\, dy \equiv 
 \sqrt{|\mathrm{det}(a_{\alpha\beta})|} \,\dd y\,.
\end{equation}
The second and third fundamental forms $b_{\alpha\beta}$, $c_{\alpha\beta}$ are given by
\begin{subequations}
\begin{eqnarray}\label{eq:fundamental_forms}
b_{\alpha\beta} &=& \bs n \cdot \bs a_{\alpha,\beta}\,,\\
c_{\alpha\beta} &=& \bs n_{,\alpha} \cdot \bs n_{,\beta}\,.
\end{eqnarray}
\end{subequations}
The covariant derivative is denoted by $\nabla_{\alpha}$ whereas $\nabla$ is the surface gradient. For a scalar field $\psi$, its components are $\nabla^{\alpha} \psi = a^{\alpha \beta} \nabla_{\alpha} \psi$. The Laplace-Beltrami operator $\triangle = \nabla^{\alpha} \nabla_{\alpha}$ applied to a scalar function is given by
\begin{align}\label{eq:LaplaceBeltrami}
\triangle \psi  = a^{\gamma\delta} \psi_{,\gamma\delta} -a^{\gamma\delta} \Gamma^{\lambda}_{\gamma\delta} \psi_{,\lambda}\,,
\end{align}
where $\Gamma^{\lambda}_{\gamma\delta}$ are the Christoffel symbols. Explicit expressions for various differential operators applied to higher order tensors can be found e.g. in~\cite{Deserno2004}.

\section*{Curvature Potential}

The Koiter shell equations derive from the (dimensionless) Koiter shell energy~\cite{Niordson1985}
\begin{equation}
\begin{aligned}
\overline{\mathcal{U}}^\textup{K}&=\overbrace{\int [(1-\nu)\tr(\vett{a}-\accentset{\circ}{\vett{a}})^2+\nu\tr^2(\vett{a}-\accentset{\circ}{\vett{a}})]\,\dd\omega}^{\overline{\mathcal{U}}_\textup{s}^\textup{K}}\\&+\frac{h^2}{3}\underbrace{\int [(1-\nu)\tr(\vett{b}-\accentset{\circ}{\vett{b}})^2+\nu\tr^2(\vett{b}-\accentset{\circ}{\vett{b}})]\,\dd\omega}_{\overline{\mathcal{U}}_\textup{b}^\textup{K}}\,,
\end{aligned}
\end{equation}
where~$\accentset{\circ}{\vett{a}}$ and~$\accentset{\circ}{\vett{b}}$ are the first and second fundamental forms of the undeformed shell, $\nu$ is the Poisson ratio, $\tr$ denotes the trace operator in the surface metric, $\dd\omega$ is the area element defined by~$\dd\omega=\sqrt{\det\accentset{\circ}{\vett{a}}}\ \dd\eta^1\dd\eta^2$, $\vett{a}$ is the first fundamental form of the mid-surface, containing all information about lateral distances between points, and~$\vett{b}$ is the second fundamental form of the mid-surface, containing all information about the local curvature. The energy of a non-Euclidean shell~$\overline{\mathcal{U}}$ can be obtained by the Koiter energy by substituting~$\vett{\bar{a}}$ and~$\vett{\bar{b}}$ for~$\accentset{\circ}{\vett{a}}$ and~$\accentset{\circ}{\vett{b}}$, respectively, where the \textit{natural} first and second fundamental forms~$\vett{\bar{a}}$ and~$\vett{\bar{b}}$ represent the lateral distances and curvatures that would make the sheet locally stress-free, and they are determined by the specific stimulus~\cite{Efrati2009,Pezzulla2017}. This also introduces the conformal stretching factor~$\Lambda_\textup{o}$ in front of the bending energy, and a consistent change of all the metric-defined operators.
In this paper, we focus on those stimuli that induce a natural curvature without a change of the first fundamental form. This can be stated as~$\bar{\vett{a}}=\accentset{\circ}{\vett{a}}$ ($\Lambda_\textup{o}=1$) and~$\bar{\vett{b}}=\accentset{\circ}{\vett{b}}+\kappa\accentset{\circ}{\vett{a}}$, where~$\accentset{\circ}{\vett{a}}$ and~$\accentset{\circ}{\vett{b}}$ are the first and second fundamental forms of the undeformed shell, and $\kappa$ is the (additive) natural curvature of the shell.
%The energy of a non-Euclidean shell is usually written as~\cite{Efrati2009,Pezzulla2017}
%
%\begin{equation}
%\begin{aligned}
%\overline{\mathcal{U}}&=\overbrace{\int [(1-\nu)\tr(\vett{a}-\vett{\bar{a}})^2+\nu\tr^2(\vett{a}-\vett{\bar{a}})]\,\dd\bar{\omega}}^{\overline{\mathcal{U}}_\textup{s}}\\&+\Lambda_\textup{o}^2\frac{h^2}{3}\underbrace{\int [(1-\nu)\tr(\vett{b}-\vett{\bar{b}})^2+\nu\tr^2(\vett{b}-\vett{\bar{b}})]\,\dd\bar{\omega}}_{\overline{\mathcal{U}}_\textup{b}}\,,
%\end{aligned}
%\end{equation}
Let us notice that
\begin{equation}
\begin{aligned}
\tr(\vett{b}-\vett{\bar{b}})^2&=\tr(\vett{b}-\accentset{\circ}{\vett{b}})^2-2\kappa\, \tr(\vett{b}-\accentset{\circ}{\vett{b}})+2\kappa^2\,,\\
\tr^2(\vett{b}-\vett{\bar{b}})&=\tr^2(\vett{b}-\accentset{\circ}{\vett{b}})-4\kappa\, \tr(\vett{b}-\accentset{\circ}{\vett{b}})+4\kappa^2\,,\\
\end{aligned}
\end{equation}
which are coordinate-free expressions, independent of the shape of the shell. The energy of non-Euclidean shells may be then written as

\begin{equation}\label{Senergy}
\overline{\mathcal{U}}=\overline{\mathcal{U}}_{\textup{s}}^{\textup{K}}+\frac{h^2}{3}\overline{\mathcal{U}}_{\textup{b}}^{\textup{K}}-\frac{2(1+\nu)}{3}h^2\int\kappa\,\tr({\vett{b}}-\accentset{\circ}{\vett{b}})\,\dd\omega\,,
\end{equation}
where we ignored the term proportional to~$\kappa^2$, which does not depend on the state variables ($\vett{a}$ and~$\vett{b}$) and therefore does not affect the minimization problem. The previous expression coincides with Eq.~(1) in the main text.

\begin{figure}[t]
\centering
\includegraphics[scale=1]{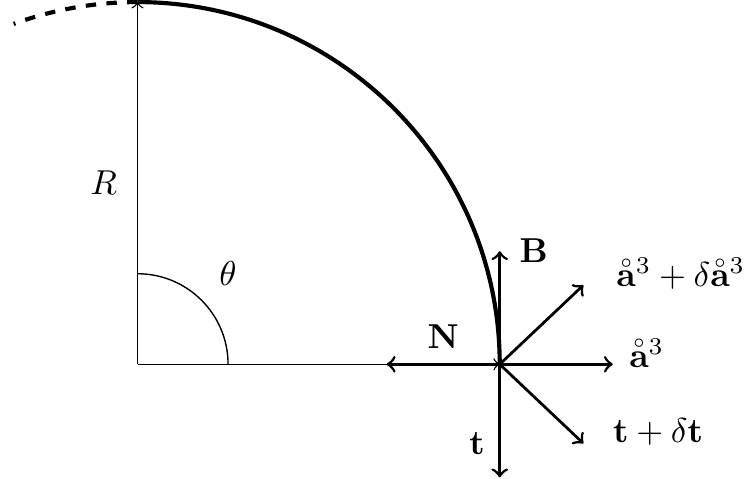}
\caption{Axisymmetric profile of the shell with variations of the normal and the unit tangent vector on the boundary curve. \label{curvatures}}
\end{figure}

We can interpret the last factor of Eq.~\eqref{Senergy} as the potential~$\mathcal{P}_\kappa$ of the spontaneous curvature. If we denote the displacement field by~$\vett{\Psi}=\Psi_i\accentset{\circ}{\vett{a}}^i$, we have~\cite{Niordson1985}

\begin{equation}\label{traceb}
\tr(\vett{b}-{\accentset{\circ}{\vett{b}}})=\triangle\Psi_3-\accentset{\circ}{c}_\alpha^\alpha\Psi_3+2\accentset{\circ}{b}^{\beta\sigma}\nabla_\beta{\Psi_\sigma}+\nabla^\alpha \accentset{\circ}{b}_\alpha^\tau \Psi_\tau+O(|\vett{\Psi}|^2)\,,
\end{equation}
where~$\accentset{\circ}{\vett{c}}$ is the third fundamental form of the undeformed surface. The trace of the third fundamental form is~$\tr\accentset{\circ}{\vett{c}}=\accentset{\circ}{c}_\alpha^\alpha=4\accentset{\circ}{\mathcal{H}}^2-2\accentset{\circ}{\mathcal{K}}$, where~$\accentset{\circ}{\mathcal{H}}$ and~$\accentset{\circ}{\mathcal{K}}$ are the mean and the Gaussian curvatures of the undeformed mid-surface, respectively~\cite{Deserno2004}. For a homogeneous~$\kappa$, if we substitute Eq.~\eqref{traceb} into Eq.~\eqref{Senergy} and use the Leibniz rule, we get for the curvature potential
\begin{equation}
\begin{aligned}
\mathcal{P}_\kappa&=-\frac{2(1+\nu)}{3}h^2\kappa\int[\triangle\Psi_3-\accentset{\circ}{c}_\alpha^\alpha\Psi_3+\nabla_\beta(\accentset{\circ}{b}^{\beta\sigma}{\Psi_\sigma})\\&+\accentset{\circ}{b}^{\beta\sigma}\nabla_\beta\Psi_\sigma]\,\dd\omega\,.
\end{aligned}
\end{equation}
By applying the generalized Stokes theorem to the terms in divergence form, we obtain generalized bulk and boundary work terms:
\begin{equation}
\begin{aligned}
&\mathcal{W}_\textup{edge}=\frac{2(1+\nu)}{3}h^2\kappa\oint\vett{q}\cdot\vett{t}\,\dd s\,,\\
&\mathcal{W}_\textup{bulk}=\frac{2(1+\nu)}{3}h^2\kappa\int\accentset{\circ}{b}^{\beta\sigma}\nabla_\beta\Psi_\sigma-\accentset{\circ}{c}_\alpha^\alpha\Psi_3\,\dd \omega\,,\\
\end{aligned}
\end{equation}
where $\cdot$ denotes the inner product in the surface metric, $\vett{t}$ is the unit tangent vector to the surface, perpendicular to the boundary curve, and~$\vett{q}$ is the rotation vector. For small displacements ($=O(h)$), the rotation of a covariant vector~$\accentset{\circ}{\vett{a}}_\alpha$ is~$q_\alpha=\delta\accentset{\circ}{\vett{a}}_\alpha\cdot\accentset{\circ}{\vett{a}}^3$, where~$\delta$ denotes a small increment (Figure~\ref{curvatures}), therefore one can also define the rotation vector as
\begin{equation}
\vett{q}=(\delta\accentset{\circ}{\vett{a}}_\alpha\cdot\accentset{\circ}{\vett{a}}^3)\accentset{\circ}{\vett{a}}^\alpha=-(\delta\accentset{\circ}{\vett{a}}^3\cdot\accentset{\circ}{\vett{a}}_\alpha)\accentset{\circ}{\vett{a}}^\alpha=\nabla\Psi_3+\accentset{\circ}{\vett{b}}\check{\vett{\Psi}}\,.
\end{equation}
By introducing the characteristic length of the boundary curve~$\bar{l}$, radius of curvature~$\bar{R}$ and area of the shell~$\bar{A}$, we can generalize the scalings presented in the main text, and show that~$\mathcal{W}_\textup{edge}\sim h^2\kappa^2\bar{l}\sqrt{\bar{R}h}$ and~$\mathcal{W}_\textup{bulk}\sim h^4\kappa^2\bar{R}^{-2}\bar{A}$, where we have used that for near isometries, $|\check{\vett{\Psi}}|<\Psi_3$. The scalings of the generalized work terms suggest that the boundary work prevails over the bulk work for open shells, regardless of the shape of the mid-surface.
For surfaces with~$\nabla^\alpha\accentset{\circ}{b}_\alpha^\tau=0$, such as plates, cylinders, and spheres, among others, the work terms can be reworked as
\begin{equation}
\begin{aligned}
&\mathcal{W}_\textup{edge}=\frac{2(1+\nu)}{3}h^2\kappa\oint(\vett{q}+\accentset{\circ}{\vett{b}}\check{\vett{\Psi}})\cdot\vett{t}\,\dd s\,,\\
&\mathcal{W}_\textup{bulk}=-\frac{2(1+\nu)}{3}h^2\kappa\int\accentset{\circ}{c}_\alpha^\alpha\Psi_3\,\dd \omega\,,
\end{aligned}
\end{equation}
where now the boundary work suggests a combined torque and membrane force at the boundary, while the bulk work suggests an effective pressure in the bulk.

%For a sphere, the two work terms become equations~(2) in the main text:
%\begin{subequations}
%\begin{eqnarray}
%\mathcal{W}_\textup{bulk}&=&-\frac{4(1+\nu)}{3}\Bigl(\frac{h}{R}\Bigr)^2\kappa\int \Psi_3\,\dd\omega\,,\\
%\mathcal{W}_\textup{edge}&=&\frac{2(1+\nu)}{3}h^2\kappa\oint\Bigl(\vett{q}-\frac{\check{\vett{\Psi}}}{R}\Bigr)\cdot\vett{t}\,\dd s\,.
%\end{eqnarray}
%\end{subequations}

\subsection{Scaling of $|\vett{q}-\check{\vett{\Psi}}/R|$}

As the boundary layer is bending dominated, we have~$\tr(\vett{b}-\accentset{\circ}{\vett{b}})\sim\kappa$. Since~$\tr(\vett{b}-\accentset{\circ}{\vett{b}})=\nabla^\alpha(\nabla_\alpha\Psi_3-2\Psi_\alpha/R)-2\Psi_3/R^2\sim|\vett{q}-\check{\vett{\Psi}}/R|/\sqrt{Rh}$, we conclude that~$|\vett{q}-\check{\vett{\Psi}}/R|\sim\kappa\sqrt{Rh}$.

\section*{Snapping of Open Shells}

\begin{figure}[t]
\centering
\includegraphics[scale=1.1]{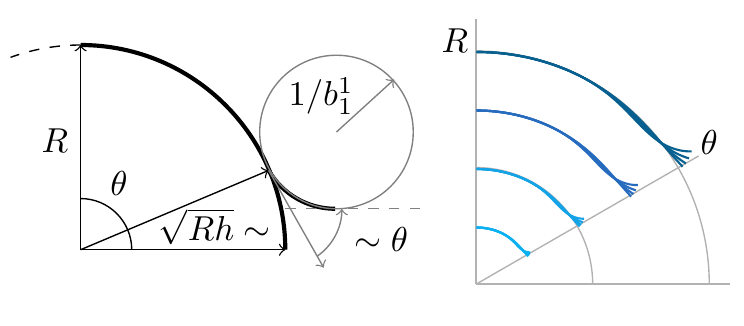}
\caption{A spherical shell at snapping, the observation that its tangent plane on the boundary curve is horizontal suggests the scaling analysis (left). Self-similar profiles of shells before snap-through instabilities (right).\label{scaling}}
\end{figure}

\subsection{Scaling}

\begin{figure} 
\centering
\includegraphics[scale=1]{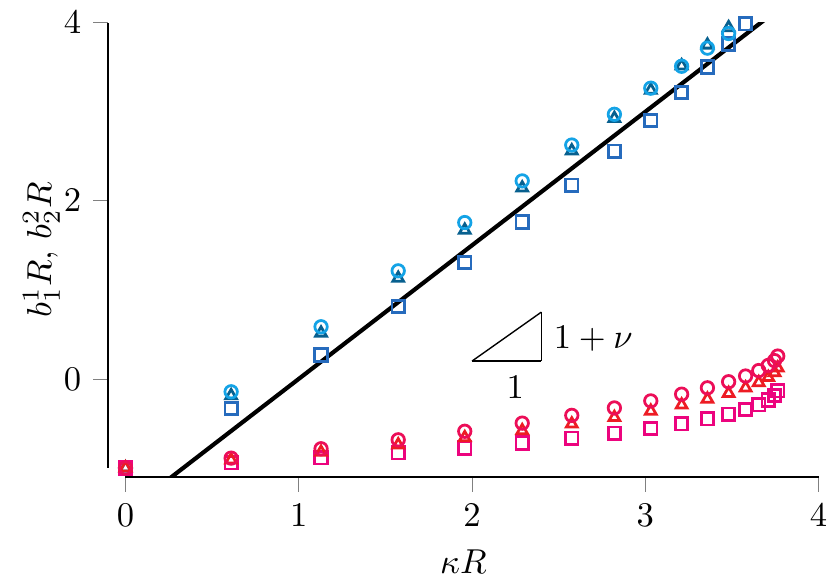}
\caption{Longitudinal ($b_1^1$, blue symbols) and meridional ($b_2^2$, red symbols) curvature within the boundary layer before snapping, at different locations and for different geometries, for different values of~$h/R\in[0.001,0.1]$. The solid black line represents Eq.~\eqref{b11} whereas symbols represent numerical simulations. \label{curvBL}}
\end{figure}

The experimental observation that shells with~$\theta<\pi/2$ snap when the tangent plane on the boundary curve is horizontal leads to the scaling law represented by Eq. (3) in the main text, which we derive here in more detail with reference to Figure~\ref{scaling}. At snapping, the tangent plane on the boundary curve is horizontal meaning that the covariant vector~$\vett{a}_1$ is horizontal. If we want to estimate the characteristic curvature within the boundary layer with a width scaling as~$\sqrt{Rh}$, we have to estimate the characteristic variation of the angle between~$\vett{a}_1$ and the horizontal axis. Indeed, $\vett{a}_1$ rotates by an amount proportional to~$\theta$ from the beginning of the boundary layer up to the boundary curve. Consequently, we can estimate the characteristic curvature as~$\theta/\sqrt{Rh}$. At the same time, we can estimate this curvature via a minimization of the energy in this state, assuming that it is bending-dominated~\cite{Efrati2009a} and almost flat ($|b_2^2|\ll|b_1^1|$). With these two assumptions, the Euler-Lagrange equations associated with the bending energy read
\begin{equation}
\begin{aligned}
&2\Bigl[b_1^1-\Bigl(-\frac{1}{R}+\kappa\Bigr)\Bigr]+2\nu\Bigl[b_2^2-\Bigl(-\frac{1}{R}+\kappa\Bigr)\Bigr]-\lambda b_2^2=0\,,\\
&2\nu\Bigl[b_1^1-\Bigl(-\frac{1}{R}+\kappa\Bigr)\Bigr]+2\Bigl[b_2^2-\Bigl(-\frac{1}{R}+\kappa\Bigr)\Bigr]-\lambda b_1^1=0\,,\\
\end{aligned}
\end{equation} 
where~$\lambda$ is the Lagrange multiplier enforcing the constraint of a null Gaussian curvature, $b_1^1$ and~$b_2^2$ are the curvatures along the colatitude and azimuthal directions in the boundary layer, respectively. Using the assumption~$b_2^2\simeq0$, we get~$\lambda=2(\nu-1)$ and
\begin{equation}\label{b11}
b_1^1\simeq(1+\nu)(-1/R+\kappa_\textup{s})\,,
\end{equation}
which nicely agrees with numerical results as shown in figure~\ref{curvBL}.
If we combine the two scalings we get Eq. (3) in the main text
\begin{equation}
(1+\nu)(-1/R+\kappa_\textup{s})\sim\frac{\theta}{\sqrt{Rh}}\Rightarrow\kappa_\textup{s}R=\beta\bar{\theta}-\alpha\,,
\end{equation}
where~$\bar{\theta}=\theta/\sqrt{h/R}$, $\nu=1/2$, and $\alpha$ and~$\beta$ are two unknown parameters that have to be determined numerically.

\subsection{1D model and 2D model}

As the deformation up to the snap-through instability is experimentally observed to be axially symmetric, we derive a 1D model in which the energy of the shell is minimized in the profile curve of the shape. We write the parametrization of the profile curve as~$\eta^1\mapsto(\phi(\eta^1),\psi(\eta^1))$, such that when~$\kappa=0$ we have~$\phi(\eta^1)=R\sin\eta^1$ and~$\psi(\eta^1)=R\cos\eta^1$ with~$\eta^1\in[0,\theta]$. We then derive the first and second fundamental forms as~\cite{Efrati2009}
\begin{equation}
\begin{aligned}
&\vett{a}=
\begin{pmatrix}
\phi_u^2+\psi_u^2 &0\\
0&\phi^2\\
\end{pmatrix}\,,\\
&\vett{b}=\frac{1}{\sqrt{\phi_u^2+\psi_u^2}}
\begin{pmatrix}
\psi_{uu}\phi_u-\phi_{uu}\psi_u&0\\
0&\phi\psi_u\\
\end{pmatrix}\,.
\end{aligned}
\end{equation}
If we substitute these expressions in the energy of the shell, we obtain a second order functional of~$\phi$ and~$\psi$, and we reduce it to a first order functional with the change of variable~$\Psi_u=\phi_u\zeta$~\cite{Efrati2009}. For symmetry we require that~$\phi(0)=\zeta(0)=0$ and~$\psi(0)=R$, $\forall\kappa$.
We minimize the first order functional in COMSOL Multiphysics with a custom arc-length method to vary~$\kappa$.

The numerical results confirm the scaling law~$\kappa_\textup{s}R=\beta\bar{\theta}-\alpha$, with~$\alpha=0.67$ and~$\beta=0.85$. The prominent role of geometry throughout this analysis leads to the self-similarity shown in Figure~\ref{scaling}~(right), where shells possessing the same values of~$h/R$, $\theta$, and~$\kappa R$ have similar shapes that collapse onto a single one upon the affine transformation of uniform scaling by~$1/R$.

In the~$2$D model, we minimize the energy of the shell via a C1-continuous subdivision finite elements (SDFEs)~\cite{Cirak2001}.

\section*{Buckling of Open Shells}

\begin{figure} 
\centering
\includegraphics[scale=1]{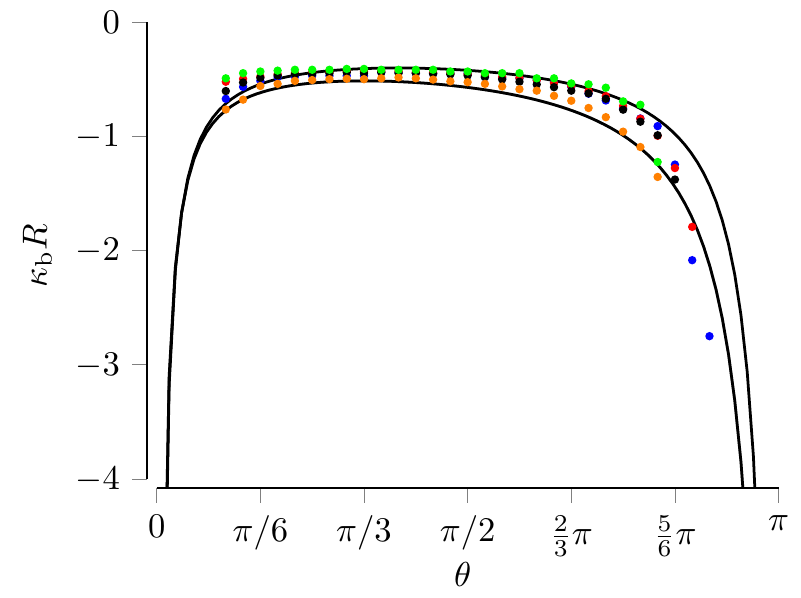}
\caption{Critical buckling curvature versus~$\theta$. Circles in green, red, blue, black, and orange denote 2D simulations for~$h/R=0.01,0.018,0.02,0.022,0.04$, respectively, while black solid lines represent Eq.~(4). \label{sphere}}
\end{figure}

Let us now focus on the buckling of open shells with particular emphasis on the two diverging regimes. For~$\theta\rightarrow0$ we derived
$
\kappa_\textup{b}R=\pm a/\bar{\theta}^2+1\,,
$
from the flat case presented in~\cite{Pezzulla2016}, assuming the radius of the plate to be equal to~$R\theta$. When~$\theta\rightarrow\pi$, we estimated a divergence of the critical curvature as~$1/(\theta-\pi)$. We represent the critical buckling curvature as a function of~$\theta$ in Figure~\ref{sphere}, where symbols denote 2D simulations, and colors denote different values of~$h/R$.
The diverging behavior for~$\theta\rightarrow0$ is very well captured by our scaling and is universal as well as the divergence for~$\theta\rightarrow\pi$, although the latter does not rely on an analytically solved plate problem. The scalar coefficients in Eq.~(4) in the main text are determined numerically as~$b\simeq3.6$ and~$c\simeq-0.98$.

\section*{Buckling of Closed Shells}

\subsection*{Fundamental solution}

Deformations preserving the spherical symmetry satisfy~$\Psi_1=\Psi_2=0$. Therefore, we can express the linear stretching and bending strains as~\cite{Niordson1985}
\begin{equation}
\begin{aligned}
\tr^2(\vett{a}-\accentset{\circ}{\vett{a}})&=\frac{16}{R^2}\Psi_3^2\,,\quad \tr(\vett{a}-\accentset{\circ}{\vett{a}})^2=\frac{8}{R^2}\Psi_3^2\,,\\
\tr^2(\vett{b}-\accentset{\circ}{\vett{b}})&=\frac{4}{R^4}\Psi_3^2\,,\quad\tr(\vett{b}-\accentset{\circ}{\vett{b}})^2=\frac{2}{R^4}\Psi_3^2\,,\\
\end{aligned}
\end{equation}
so that the Euler-Lagrange equation for~$\Psi_3$ is
\begin{equation}
\frac{4}{R^2}(1+\nu)\Bigl(4+\frac{1}{3}\Bigl(\frac{h}{R}\Bigr)^2\Bigr)\Psi_3+\frac{4}{3}(1+\nu)\Bigl(\frac{h}{R}\Bigr)^2\kappa=0\,,
\end{equation}
which gives the result of the main text
\begin{equation}\label{Psi3_fund}
\Psi_3=-\frac{R^2}{12}\frac{(h/R)^2}{1+(1/12)(h/R)^2}\kappa=-\frac{h^2}{12}\kappa+O((h/R)^4)\,.
\end{equation}
This expression is consistent with that obtained in the case of pressure buckling~\cite{Hutchinson1967} via the analogy presented via Eq.~(5) in the main text. Indeed, if we substitute Eq.~(5) in Eq.~\eqref{Psi3_fund}, we find
\begin{equation}
\Psi_3=-(1-\nu)\frac{pR^2}{2Eh}\,.
\end{equation}
This configuration of the shell is called \emph{fundamental state}.

\subsection*{Stability analysis}

While the analogy between spontaneous curvature and pressure allows for a better understanding of the spontaneous curvature and for a simple determination of the fundamental state, it does not adequately describe the residual stress that develops within the shell. Indeed, while the residual stress developed within a pressurized thin shell is of the membrane type (residual bending moments are negligible), a spontaneous curvature modifies the \emph{bending} ground state leading to residual bending moments that have the same order of magnitude of the residual membrane stress.

To determine the critical spontaneous curvature leading to buckling, we have to enrich the procedure presented by Koiter~\cite{heijden2008} to take into account the residual bending moments. We therefore write the quadratic energy functional as
\begin{equation}\label{func}
\begin{aligned}
P_2[\vett{\Psi};\kappa]&=\int \frac{Eh}{2(1-\nu^2)}\Bigl[(1-\nu)G^{\alpha\beta}G_{\alpha\beta}+\nu(G_\alpha^\alpha)^2\\
&+\frac{h^2}{12}[(1-\nu)\rho^{\alpha\beta}\rho_{\alpha\beta}+\nu(\rho_\alpha^\alpha)^2]\Bigr]+N^{\alpha\beta}\gamma_{\alpha\beta}\\
&+M^{\alpha\beta}\zeta_{\alpha\beta}\ \dd\omega\,,
\end{aligned}
\end{equation}
where~$G_{\alpha\beta}$ and~$\rho_{\alpha\beta}$ represent the linear stretching and bending strains with respect to the fundamental state, $\vett{\Psi}$~is here the displacement field from the fundamental state, $\gamma_{\alpha\beta}$ and~$\zeta_{\alpha\beta}$ are the nonlinear second order stretching and bending strains with respect to the fundamental state. The term in the last line represents the contribution of the bending moments in the fundamental state to the energy functional, which is neglected in the classical pressure buckling problem because of the absence of inelastic stimuli that change the ground state.

We now report the expressions for the linear and nonlinear (second order) strains, and evaluate~$M^{\alpha\beta}\zeta_{\alpha\beta}$. Linear stretching and bending strains can be written as~\cite{Niordson1985}
\begin{equation}\label{lin_strains}
\begin{aligned}
G_{\alpha\beta}&=\frac{1}{2}(\nabla_\beta\Psi_\alpha+\nabla_\alpha\Psi_\beta)+\frac{\Psi_3}{R}\accentset{\circ}{a}_{\alpha\beta}\,,\\
\rho_{\alpha\beta}&=\nabla_{\alpha\beta}\Psi_3-\frac{1}{R^2}\accentset{\circ}{a}_{\alpha\beta}\Psi_3-\frac{1}{R}(\nabla_\beta\Psi_\alpha+\nabla_\alpha\Psi_\beta)\,,
\end{aligned}
\end{equation}
where the first and second fundamental forms in the fundamental state can be approximated with those in the undeformed state, since the displacement from the undeformed to the fundamental state is smaller than the thickness~\cite{heijden2008}. The nonlinear (second order) stretching strain is
\begin{equation}
\begin{aligned}
\gamma_{\alpha\beta}&=\frac{1}{2}\nabla^\lambda\Psi_\alpha\nabla_\lambda\Psi_\beta+\frac{1}{2R}\Psi_3\nabla_\beta\Psi_\alpha+\frac{1}{2R}\Psi_3\nabla_\alpha\Psi_\beta\\
&+\frac{1}{2R^2}\accentset{\circ}{a}_{\alpha\beta}\Psi_3^2+\frac{1}{2}\nabla_\alpha\Psi_3\nabla_\beta\Psi_3-\frac{1}{R}\Psi_\alpha\nabla_\beta\Psi_3\\
&+\frac{1}{2R^2}\Psi_\alpha\Psi_\beta\,.
\end{aligned}
\end{equation}
As regards the nonlinear (second order) bending strain, the expression is rather lengthy. However, we shall see that only their trace is needed for the analysis. Indeed, we compute the bending moments in the fundamental state as
\begin{equation}
M^{\alpha\beta}=\frac{Eh^3}{12(1+\nu)}\Bigl(\frac{\nu}{1-\nu}\accentset{\circ}{a}^{\alpha\beta}\accentset{\circ}{a}^{\gamma\delta}+\accentset{\circ}{a}^{\alpha\gamma}\accentset{\circ}{a}^{\beta\delta}\Bigr)(b_{\gamma\delta}|_{f}-\bar{b}_{\gamma\delta})\,,
\end{equation}
where~$b_{\gamma\delta}|_{f}$ denotes the second fundamental form in the fundamental state. Combining Eq.~\eqref{Psi3_fund} with strains~\eqref{lin_strains} arising from a displacement from the undeformed to the fundamental state, we get
\begin{equation}
b_{\gamma\delta}|_{f}-\bar{b}_{\gamma\delta}=-\Bigl(-\frac{1}{12}\Bigl(\frac{h}{R}\Bigr)^2+1\Bigr)\kappa\accentset{\circ}{a}_{\gamma\delta}\simeq-\kappa\accentset{\circ}{a}_{\gamma\delta}\,,
\end{equation}
where the last approximation is consistent within a theory for thin shells. Therefore, the bending moments in the fundamental state are
\begin{equation}
M^{\alpha\beta}=-\frac{Eh^3}{12(1-\nu)}\kappa\accentset{\circ}{a}^{\alpha\beta}\,,
\end{equation}
and its contraction in~\eqref{func} with the bending strains therefore amounts to taking the trace of~$\zeta_{\alpha\beta}$. From~\cite{Deserno2004}, for a spherical surface we have
\begin{equation}\label{bending}
\begin{aligned}
\accentset{\circ}{a}^{\alpha\beta}\zeta_{\alpha\beta}&=\frac{1}{R^2}\Psi^\alpha\nabla_\alpha\Psi_3-\frac{1}{R}|\nabla\Psi_3|^2\\
&+\frac{1}{R}\Psi_\alpha\triangle\Psi^\alpha-\nabla_\alpha\Psi_3\triangle\Psi^\alpha\,.
\end{aligned}
\end{equation}
In order to express~$M^{\alpha\beta}\zeta_{\alpha\beta}$ in a convenient way, we recall that the tangential displacement can always be expressed in terms of two invariants~\cite{Neut1932,heijden2008}
\begin{equation}
\Psi_\alpha=\nabla_\alpha\phi+\varepsilon_{\alpha\lambda}\nabla^\lambda\psi\,,
\end{equation}
where~$\phi$ and~$\psi$ can be regarded as potentials, and~$\varepsilon_{\alpha\lambda}$ is the two-dimensional Levi-Civita symbol. The advantage of writing the in-plane displacement via the two potentials fields is that the balance equations will be decoupled in~$\phi$ and~$\psi$.

Let us start with the first term in Eq.~\eqref{bending}:
\begin{equation}
\begin{aligned}
\frac{1}{R^2}\int\Psi^\alpha\nabla_\alpha\Psi_3\, \dd\omega&=\frac{1}{R^2}\int\nabla_\alpha(\Psi^\alpha\Psi_3)-\triangle\phi\Psi_3\, \dd\omega=\\
&=-\frac{1}{R^2}\int\triangle\phi\Psi_3\,\dd\omega\,,
\end{aligned}
\end{equation}
where we used the generalized Stokes theorem on a closed surface and~$\nabla_\alpha\Psi^\alpha=\triangle\phi$, since~$\nabla_{\alpha\beta}\psi$ is symmetric. As regards the second term in Eq.~\eqref{bending}
\begin{equation}
\begin{aligned}
-\frac{1}{R}\int|\nabla\Psi_3|^2\,\dd\omega&=-\frac{1}{R}\int\nabla_\alpha(\nabla^\alpha\Psi_3)-\Psi_3\triangle\Psi_3\,\dd\omega=\\
&=\frac{1}{R}\int\Psi_3\triangle\Psi_3\,\dd\omega\,,
\end{aligned}
\end{equation}
again by using the generalized Stokes theorem on a closed surface. For the third term in Eq.~\eqref{bending} we have
\begin{equation}\label{psiapsia}
\begin{aligned}
\frac{1}{R}\int\Psi_\alpha\triangle\Psi^\alpha\,\dd\omega&=\frac{1}{R}\int\nabla_\alpha\phi\nabla^{\alpha\beta}_{\cdot\cdot\beta}\phi+\varepsilon_{\alpha\lambda}\nabla^\lambda\psi\nabla^{\alpha\beta}_{\cdot\cdot\beta}\phi\\
&+\nabla_\alpha\phi\varepsilon^\alpha_\gamma\nabla^{\gamma\cdot\beta}_{\cdot\beta}\psi+\varepsilon_{\alpha\gamma}\nabla^\gamma\psi\varepsilon^\alpha_\lambda\nabla^{\lambda\beta}_{\cdot\cdot\beta}\psi\,\dd\omega\,.
\end{aligned}
\end{equation}
Following the procedure in~\cite{heijden2008}, the second and third in~\eqref{psiapsia} terms are zero, whereas the first term can be simplified as
\begin{equation}
\frac{1}{R}\int\nabla_\alpha\phi\nabla^{\alpha\beta}_{\cdot\cdot\beta}\phi\,\dd\omega=-\frac{1}{R}\int\Bigl((\triangle\phi)^2+\frac{\triangle\phi}{R^2}\phi\Bigr)\,\dd\omega\,.
\end{equation}
Analogously, the fourth term can be rewritten as
\begin{equation}
\frac{1}{R}\int\varepsilon_{\alpha\gamma}\nabla^\gamma\psi\varepsilon^\alpha_\lambda\nabla^{\lambda\beta}_{\cdot\cdot\beta}\psi\,\dd\omega=-\frac{1}{R}\int\Bigl((\triangle\psi)^2+\frac{\triangle\psi}{R^2}\psi\Bigr)\,\dd\omega\,,
\end{equation}
by applying the generalized Stokes theorem. Finally, for the fourth term in Eq.~\eqref{bending} we find
\begin{equation}
\begin{aligned}
-\int\nabla_\alpha\Psi_3\triangle\Psi^\alpha\,\dd\omega&=\int\nabla_{\alpha\beta}\Psi_3\nabla^{\alpha\beta}\phi\,\dd\omega=\\
&=\int\triangle\phi\Bigl(\triangle\Psi_3+\frac{\Psi_3}{R^2}\Bigr)\,\dd\omega\,.
\end{aligned}
\end{equation}
Using these simplifications, the contribution of the bending moments in the fundamental state becomes
\begin{equation}\label{moments}
\begin{aligned}
\int M^{\alpha\beta}\zeta_{\alpha\beta}\,\dd\omega&=-\frac{Eh^2}{12(1-\nu)}\kappa h\int\frac{1}{R}\Psi_3\triangle\Psi_3+\triangle\phi\triangle\Psi_3\\
&-\frac{(\triangle\phi)^2}{R}-\frac{\triangle\phi}{R^3}\phi-\frac{(\triangle\psi)^2}{R}-\frac{\triangle\psi}{R^3}\psi\,\dd\omega\,.
\end{aligned}
\end{equation}
We shall show that the terms in the second line can be neglected. To do so, we express Koiter's result on the contribution of the membrane stress in the fundamental state in the case of a natural curvature via the analogy presented in the main text as
\begin{equation}\label{membrane}
\begin{aligned}
\int N^{\alpha\beta}\gamma_{\alpha\beta}\,\dd\omega&=-\frac{Eh^2}{24(1-\nu)R}\kappa h\int (\triangle\phi)^2+4\frac{\Psi_3}{R}\triangle\phi\\
&-\Psi_3\triangle\Psi_3+\frac{2}{R^2}\Psi_3^2+(\triangle\psi)^2\,\dd\omega\,,
\end{aligned}
\end{equation}
where we also took into account the different convention used by Koiter about the orientation of the normal. We recall now that Koiter neglected all terms in the last functional except~$\Psi_3\triangle\Psi_3$ by comparing them with the similar ones in the elastic energy and showing that they are smaller by at least a factor~$h/R$. Now, since the terms proportional to~$(\triangle\phi)^2$ and~$(\triangle\psi)^2$ in Eq.~\eqref{moments} have smaller pre-factors than those in Eq.~\eqref{membrane}, which Koiter neglected, can be neglected. Moreover, Koiter also neglected the following term arising from the membrane prestress
\[
\frac{Eh}{\sqrt{3(1-\nu^2)}R^2}\frac{h}{R}\int \phi\triangle\phi+\psi\triangle\psi\,\dd\omega\,,
\]
which are comparable to the similar terms in Eq.~\eqref{moments}, which we can therefore neglect. Finally, the contribution of the bending moments in the fundamental state has been reduced to
\begin{equation}\label{moments2}
\int M^{\alpha\beta}\zeta_{\alpha\beta}\,\dd\omega=-\frac{Eh^2}{12(1-\nu)}\kappa h\int\frac{1}{R}\Psi_3\triangle\Psi_3+\triangle\phi\triangle\Psi_3\,\dd\omega\,,
\end{equation}
while the contribution of the membrane stress has been reduced to
\begin{equation}\label{membrane2}
\int N^{\alpha\beta}\gamma_{\alpha\beta}\,\dd\omega=\frac{Eh^2}{24(1-\nu)R}\kappa h\int\Psi_3\triangle\Psi_3\,\dd\omega\,.
\end{equation}
Furthermore,~$\triangle\psi=0$~\cite{heijden2008}, such that we finally obtain
\begin{equation}
\begin{aligned}
P_2[\vett{\Psi};\kappa]&=\frac{Eh}{2(1-\nu^2)}\int(\triangle\phi)^2+\frac{1-\nu}{R^2}\phi\triangle\phi\\
&+\frac{2(1+\nu)}{R}\Psi_3\triangle\phi+\frac{2(1+\nu)}{R^2}\Psi_3^2+\frac{h^2}{12}(\triangle\Psi_3)^2\\
&-\frac{1+\nu}{12}\frac{h}{R}\kappa h\Psi_3\triangle\Psi_3-\frac{(1+\nu)h^2}{6}\kappa\triangle\phi\triangle\Psi_3\,\dd\omega\,.
\end{aligned}
\end{equation}
The resulting balance equations obtained via a first variation are
\begin{equation}\label{balance}
\begin{aligned}
&\triangle^2\phi+\frac{1-\nu}{R^2}\triangle\phi+\frac{1+\nu}{R}\triangle\Psi_3-\frac{1+\nu}{12}h^2 \kappa\triangle^2\Psi_3=0\,,\\
&\frac{h^2}{12}\triangle^2\Psi_3-\frac{1+\nu}{12}\frac{h}{R}\kappa h\triangle\Psi_3+2\frac{1+\nu}{R^2}\Psi_3+\frac{1+\nu}{R}\triangle\phi\\
&\qquad\qquad\qquad\qquad\qquad\qquad\,\quad-\frac{1+\nu}{12}h^2\kappa\triangle^2\phi=0\,.
\end{aligned}
\end{equation}
The standard procedure is now to expand~$\phi$ and~$\Psi_3$ in spherical harmonics
\begin{equation}
\phi(\eta^\alpha)=R\sum_{n=0}^{\infty}D_nS_n(\eta^\alpha)\,,\quad\Psi_3(\eta^\alpha)=\sum_{n=0}^\infty C_nS_n(\eta^\alpha)\,,
\end{equation}
which allows to express the system~\eqref{balance} as
\begin{equation}\label{system}
\begin{aligned}
&-\Bigl[1+\nu+\frac{1+\nu}{12}\frac{h}{R}\kappa h x\Bigr]C_n+[x-(1-\nu)]D_n=0\,,\\
&\Bigl[\frac{1}{12}\Bigl(\frac{h}{R}\Bigr)^2x^2+\frac{1+\nu}{12}\frac{h}{R}\kappa h x+2(1+\nu)\Bigr]C_n\\
&\qquad\qquad\qquad\qquad-(1+\nu)\Bigl[x+\frac{1}{12}\frac{h}{R}\kappa h x^2\Bigr]D_n=0\,,
\end{aligned}
\end{equation}
where~$x=n(n+1)$. The system~\eqref{system} has nontrivial solutions only if the determinant vanishes. As in the classical pressure buckling, the buckling mode has a short wavelength compared to the radius of the sphere, \textit{e.g.}~$x\gg1$, and the condition for nontrivial solutions reads
\begin{equation}\label{kappacri}
\kappa h=2\frac{R}{h}\frac{-3(1+2\nu)+\sqrt{3(15+(h/R)^2x^2+12\nu)}}{(1+\nu)x}\,.
\end{equation}
\begin{figure}[t] 
\centering
\hspace{-0.5cm}
\includegraphics[scale=1]{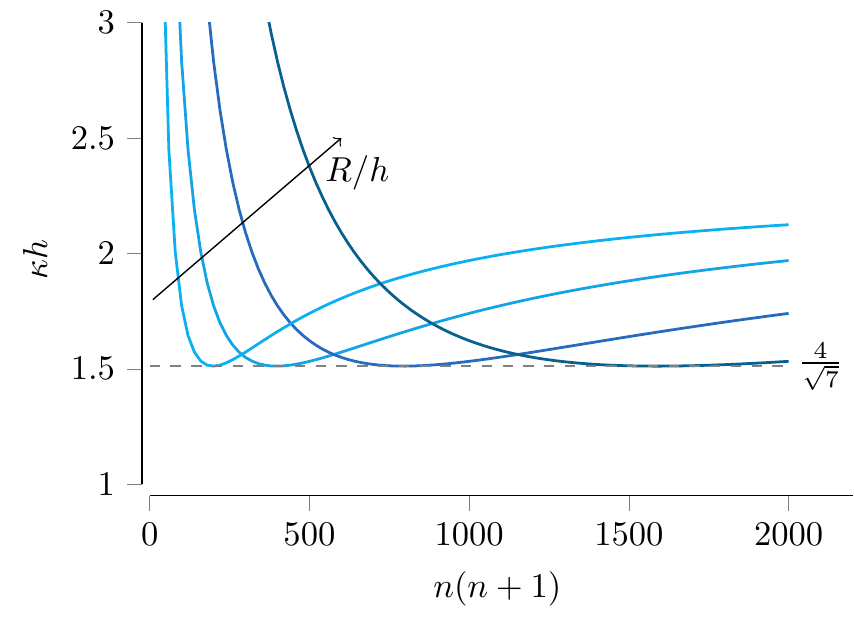}
\caption{Critical buckling curvature versus~$n(n+1)$ for~$\nu=~1/2$. \label{kappacritical}}
\end{figure}
We now look for the minimum value of the critical~$\kappa h$, which will be the buckling spontaneous curvature. To do so, as~$x\gg 1$, we regard it as a real number and minimize~\eqref{kappacri} with respect to~$x$ to get 
\begin{equation}
x=\frac{2}{1+2\nu}\frac{R}{h}\sqrt{3(5+4\nu)(1-\nu^2)}\,,
\end{equation}
which yields Eq.~(6) of the main text
\begin{equation}
\kappa_\textup{b}h=4\sqrt{3\frac{1-\nu}{(1+\nu)(5+4\nu)}}\,,
\end{equation}
which for an incompressible material becomes
\begin{equation}
\kappa_\textup{b}h=\frac{4}{\sqrt{7}}\,.
\end{equation}
Finally, as usually done in standard buckling problems in mechanics, we plot Eq.~\eqref{kappacri} as a function of~$x$ for different values of~$h/R$ in figure~\ref{kappacritical}.

\section*{Simulations on Ellipsoidal Shells}

\begin{figure}[b]
%\centering
\includegraphics[scale=0.41]{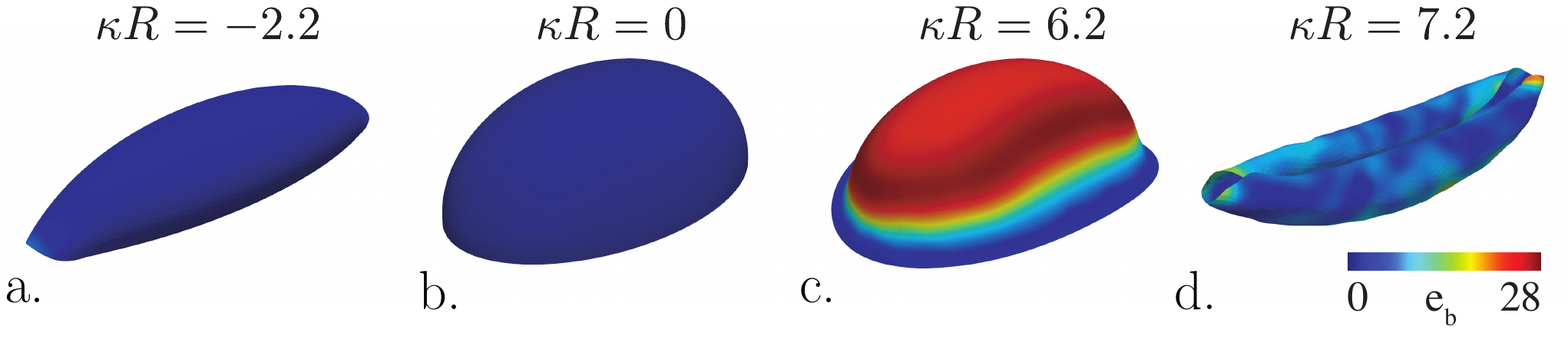}
\caption{Simulations on the curvature-induced instabilities in ellipsoidal shells. The undeformed shape is reported in \emph{b.} When the natural curvature is negative, the shell buckles into the shape shown in \emph{a.}, whereas the shell snaps when the natural curvature is positive, as reported in \emph{c.} and \emph{d.} Color denotes total bending energy density $e_b$, i.e. the last two terms of Eq.~\ref{Senergy}.\label{ellipsoids}}
\end{figure}

To show how the curvature potential can be useful to study also the morphing of shells with non homogenoeus curvatures, we performed numerical simulations on ellipsoidal shells (minor radii $r_1=r_2=R$, major radius $r_3=3/2 R$, $h/R=0.04$, opening angle $\theta = \pi/2$), in the two cases of negative and positive natural curvature. The results are reported in figure~\ref{ellipsoids}, where \emph{b.} denotes the undeformed shape. When the natural curvature is negative, the shell buckles into the shape shown in  \emph{a.}, similarly to the buckling of a spherical shell into a spindle-like shape. On the contrary, when the natural curvature is positive, the shell snaps (\emph{c.} and \emph{d.}). In particular, figure~\ref{ellipsoids} (\emph{c.}) shows that most of the deformation is focused in the boundary layer, in agreement to what the mechanical interpretation and scalings of the curvature potential for arbitrary open shells show.\\

\end{document}